\documentclass[preprint,prd,showpacs,showkeys,eqsecnum,nofootinbib]{revtex4}

\usepackage{epsfig}
\usepackage{amsmath, amssymb, amsfonts, graphics, graphicx, epsfig, epstopdf}

\begin{document}

\title{RGE of a Cold Dark Matter Two-Singlet Model}
\author{Abdessamad Abada$^{\mathrm{a,b}}$}
\email{a.abada@uaeu.ac.ae}
\author{Salah Nasri$^{\mathrm{b}}$}
\email{snasri@uaeu.ac.ae}
\affiliation{$^{\mathrm{a}}$Laboratoire de Physique des Particules et Physique
Statistique, Ecole Normale Sup\'{e}rieure, BP 92 Vieux Kouba, 16050 Alger,
Algeria\\
$^{\mathrm{b}}$Physics Department, United Arab Emirates University, POB
17551, Al Ain, United Arab Emirates}
\keywords{cold dark matter. light WIMP. extension of Standard Model. RGE.}
\pacs{95.35.+d; 98.80.-k; 12.15.-y; 11.30.Qc.}

\begin{abstract}
We study via the renormalization group equations at one-loop order the
perturbativity and vacuum stability of a two-singlet model of cold dark
matter (DM) that consists in extending the Standard Model with two real
gauge-singlet scalar fields. We then investigate the regions in the
parameter space in which the model is viable. For this, we require the model
to reproduce the observed DM relic density abundance, to comply with the
measured XENON 100 direct-detection upper bounds, and to be consistent with
the RGE perturbativity and vacuum-stability criteria up to $40\mathrm{TeV}$.
For small mixing angle $\theta $ between the physical Higgs $h$ and auxiliary
field, and DM-$h$ mutual coupling constant $\lambda _{0}^{\left( 4\right) }$,
we find that the auxiliary-field mass is confined to the interval
$116\mathrm{GeV}-138\mathrm{GeV}$ while the DM mass is mainly confined
to the region above $57\mathrm{GeV}$. Increasing $\theta $ enriches the existing
viability regions without relocating them, while increasing $\lambda
_{0}^{\left( 4\right) }$ shrinks them with a tiny relocation. We show that
the model is consistent with the recent Higgs boson-like discovery by the
ATLAS and CMS experiments, while very light dark matter
(masses below $5\mathrm{GeV}$) is ruled out by the same experiments.
\end{abstract}

\date{\today }
\maketitle

\section{Introduction}

Now that there is more and more compelling evidence that the discovery in
the ATLAS and CMS experiments at the LHC is a Higgs particle with a mass 
$m_{h}\simeq 125\mathrm{GeV}$ \cite{ATLAS, CMS}, one of the main focuses of
particle physics is an understanding of the still elusive nature of dark
matter (DM), believed to account for about 26\% of the energy content of the
universe \cite{dm-observation}. Side by side with observation, models beyond
the Standard Model (SM) are devised to account for weakly interacting
massive particles (WIMPs) as plausible candidates for dark matter. Such
models range from the sophisticated ones, those bearing intricate underlying
symmetries and mechanisms of symmetry breaking, to the more simpler ones,
those extending the SM without any particular assumption regarding a deeper
structure.

Since the findings at the LHC have so far yielded no particular clue as to
possible particle structures beyond those of the Standard Model, it is
therefore still consistent to try to recognize dark matter as simple WIMPs
extending the Standard Model, with no further assumptions as to inner
structures. The simplest of such extensions is a one real
electroweak-singlet scalar field, interacting with visible matter only via
the SM Higgs particle, first proposed by Silveira and Zee \cite{SZ} and
further studied in \cite{MD, BP, one-singlet}. In this minimal model, with
DM masses lighter than $10\mathrm{GeV}$, the Higgs would be mainly
invisible, which is excluded by the recent measurement of the Higgs signal
at the LHC \cite{Hinv}. Also, for DM masses in the range
 $7\mathrm{GeV}-60\mathrm{GeV}$, the model is ruled out by the data from XENON~10
\cite{XENON10} and CDMS II \cite{CDMSII}, except around the resonance
$62\mathrm{GeV}$ at which the Higgs-DM coupling is extremely small. Furthermore, DM\
masses between $65\mathrm{GeV}$ and $80\mathrm{GeV}$ are excluded in this
one-singlet extension by the XENON 100 experiment \cite{XENON100}. Note that
similar conclusions hold also for a complex scalar singlet extension of the
SM \cite{complex}.

So, given the difficulties this minimal model has with existing experimental
and observational DM data \cite{one-singlet-difficulties}, we have proposed
in \cite{abada-ghaffor-nasri} an extension to the Standard Model with two
real electroweak singlets, one being stable, the DM candidate, and the other
an auxiliary field with a spontaneously broken $\mathbb{Z}_{2}$ symmetry.
Based on the DM relic-density and WIMP direct detection studies, we have
concluded that this two-singlet model is capable of bearing dark matter in a
large region of the parameter space. Further constraints on the model as
well as some of its phenomenological implications have been studied in
\cite{abada-nasri-1} where rare meson decays and Higgs production channels have
been considered.

In the present work, we further the study of the two-singlet model and ask
how high in the energy scale it is computationally reliable. A\ standard
treatment is the investigation of the running of the coupling constants in
terms of the mass scale $\Lambda $ via the renorrmalization group equations
(RGE). We believe one-loop calculations are amply sufficient for the present
task; higher loops could be considered if the situation changes.

The two standard issues to monitor are the perturbativity of the scalar
coupling constants and the vacuum stability of the theory. These issues were
studied in \cite{GLR} for the complex scalar singlet extension of the SM,
and it was shown that the vacuum-stability requirement can affect the DM
relic density. Specific results from such studies depend on the cutoff scale 
$\Lambda _{m}$ of the theory. Reversely, imposing perturbativity and vacuum
stability may indicate at what $\Lambda _{m}$ the two-singlet extension is
valid. In the early parts of this work, the second point of view is adopted,
whereas in the later part, the first is taken. Furthermore, up to only
recently, it has been anticipated that new physics\ such as supersymmetry
would appear at the LHC at the scale $\Lambda \sim 1\mathrm{TeV}$. Present
results from ATLAS and CMS indicate no such signs yet. One consequence of
this is that the cutoff scale $\Lambda _{m}$ may be higher. As we shall
discuss, we find that it can be $\sim 40\mathrm{TeV}$.

This work is organized as follows. After this introduction, we recapitulate
in section II the essentials of the two-singlet model necessary for the RGE
calculations. In section III, we discuss the running of the scalar coupling
constants when we switch off the non-Higgs SM particles. This gives us a
first understanding of how high in the energy scale perturbativity is
allowed. It helps also seeing the subsequent effects of the other SM
particles. Section IV discusses the full RGEs. Vacuum stability gets into
focus with the Higgs coupling constant turning negative at some scale. The
mass scales at which non-perturbativity and non-stability set in are
different, and so a choice for $\Lambda _{m}$ has to be made. Section V
attempts at finding the regions in the parameter space in which the model is
predictive. In addition to the DM relic-density constraint and the
perturbativity-stability criteria deduced from the previous two sections, we
impose on the model to be within the current direct-detection experimental
bounds. Section VI is devoted to concluding remarks.

\section{The two-singlet model}

The model is obtained by adding to the Standard Model two real, spinless,
and $\mathbb{Z}_{2}$-symmetric SM-gauge-singlet fields. One is the dark
matter field $S_{0}$ with unbroken $\mathbb{Z}_{2}$ symmetry, and the other
an auxiliary field $\chi _{1}$ with spontaneously broken $\mathbb{Z}_{2}$
symmetry. Both fields interact with the SM particles via the Higgs doublet
$H $. Using the same notation as in \cite{abada-ghaffor-nasri}, the potential
function that involves $S_{0}$, $H$ and $\chi _{1}$ is: 
\begin{eqnarray}
U &=&\frac{\tilde{m}_{0}^{2}}{2}S_{0}^{2}-\mu ^{2}H^{\dagger }H-\frac{\mu
_{1}^{2}}{2}\chi _{1}^{2}  \notag \\
&&+\frac{\eta _{0}}{24}S_{0}^{4}+\frac{\lambda }{6}\left( H^{\dagger
}H\right) ^{2}+\frac{\eta _{1}}{24}\chi _{1}^{4}+\frac{\lambda _{0}}{2}
S_{0}^{2}H^{\dagger }H+\frac{\eta _{01}}{4}S_{0}^{2}\chi _{1}^{2}+
\frac{\lambda _{1}}{2}H^{\dagger }H\chi _{1}^{2},  \label{U}
\end{eqnarray}
where $\tilde{m}_{0}^{2}$, $\mu ^{2}$and $\mu _{1}^{2}$ and all the coupling
constants are real positive numbers\footnote{The mutual 
couplings can be negative as discussed below, see (\ref{stability-conditions}).}.

We are interested in monitoring the running of the scalar coupling
constants. A one-loop renormalization-group calculation yields the following 
$\beta $-functions:
\begin{eqnarray}
\beta _{\eta _{0}} &=&\frac{3}{16\pi ^{2}}\left( \eta _{0}^{2}+\eta
_{01}^{2}+4\lambda _{0}^{2}\right) ;  \notag \\
\beta _{\eta _{1}} &=&\frac{3}{16\pi ^{2}}\left( \eta _{1}^{2}+\eta
_{01}^{2}+4\lambda _{1}^{2}\right) ;  \notag \\
\beta _{\lambda } &=&\frac{3}{16\pi ^{2}}\left( \frac{4}{3}\lambda
^{2}+\lambda _{0}^{2}+\lambda _{1}^{2}-48\lambda _{t}^{4}+8\lambda \lambda
_{t}^{2}-3\lambda g^{2}-\lambda g^{\prime }{}^{2}+\frac{3}{2}g^{2}g^{\prime
}{}^{2}+\frac{9}{4}g^{4}\right) ;  \notag \\
\beta _{\eta _{01}} &=&\frac{1}{16\pi ^{2}}\left( 4\eta _{01}^{2}+\eta
_{0}\eta _{01}+\eta _{1}\eta _{01}+4\lambda _{0}\lambda _{1}\right) ;  \notag
\\
\beta _{\lambda _{0}} &=&\frac{1}{16\pi ^{2}}\left( 4\lambda
_{0}^{2}+\lambda _{0}\eta _{0}+2\lambda _{0}\lambda +\eta _{01}\lambda
_{1}+12\lambda _{0}\lambda _{t}^{2}-\frac{9}{2}\lambda _{0}g^{2}-\frac{3}{2}
\lambda _{0}g^{\prime }{}^{2}\right) ;  \notag \\
\beta _{\lambda _{1}} &=&\frac{1}{16\pi ^{2}}\left( 4\lambda
_{1}^{2}+\lambda _{1}\eta _{1}+2\lambda _{1}\lambda +\eta _{01}\lambda
_{0}+12\lambda _{1}\lambda _{t}^{2}-\frac{9}{2}\lambda _{1}g^{2}-\frac{3}{2}
\lambda _{1}g^{\prime }{}^{2}\right) .  \label{beta-functions-scalars}
\end{eqnarray}
As usual, $\beta _{g}\equiv dg/d\ln \Lambda $ where $\Lambda $ is the
running mass scale, starting from $\Lambda _{0}=100\mathrm{GeV}$. The
constants $g$, $g^{\prime }$ and $g_{s}$ are the SM and strong gauge
couplings, known \cite{SM-results} and given to one-loop order by the
expression:
\begin{equation}
G\left( \Lambda \right) =\frac{G\left( \Lambda _{0}\right) }
{\sqrt{1-2a_{G}\,G^{2}\left( \Lambda _{0}\right) \,\ln \left( \frac{\Lambda }
{\Lambda _{0}}\right) }},  \label{gauge-coupling-expression}
\end{equation}
where $a_{G}=\frac{-19}{96\pi ^{2}},\frac{41}{96\pi ^{2}},\frac{-7}{16\pi
^{2}}$, and $G\left( \Lambda _{0}\right) =0.65$, $0.36,1.2$ for 
$G=g,g^{\prime },g_{s}$ respectively. The coupling constant $\lambda _{t}$ is
that between the Higgs field and the top quark. To one-loop order, it runs
according to \cite{SM-results}:
\begin{equation}
\beta _{\lambda _{t}}=\frac{\lambda _{t}}{16\pi ^{2}}\left( 9\lambda
_{t}^{2}-8g_{s}^{2}-\frac{9}{4}g^{2}-\frac{17}{12}g^{\prime 2}\right) ,
\label{beta-function-top}
\end{equation}
with $\lambda _{t}\left( \Lambda _{0}\right) =\frac{m_{t}\left( \Lambda
_{0}\right) }{v}=0.7$, where $v$ is the Higgs vacuum expectation value (vev)
and $m_{t}$ the top mass. Note that we are taking into consideration the
fact that the top-quark contribution is dominant over that of the other
fermions of the Standard Model.

The model undergoes two spontaneous breakings of symmetry: one of the
electroweak, with a vev $v=246~\mathrm{GeV}$, and one of the $\mathbb{Z}_{2}$
symmetry ($\chi _{1}$ field), with a vev $v_{1}$ we take in this work equal
to $150~\mathrm{GeV}$. Above $v$, the fields and parameters of the theory
are those of (\ref{U}). Below $v_{1}$, the (scalar)\ physical fields are
$S_{0}$ (DM), $h$ (Higgs) and $S_{1}$ (auxiliary), with parameters (masses
and coupling constants) given in Eqs. (2.2--2.15) of
\cite{abada-ghaffor-nasri}. We take the values of the physical parameters at the
mass scale $\Lambda _{0}=100~\mathrm{GeV}$. There are originally nine free
physical parameters. The two vevs $v$ and $v_{1}$ are fixed, as well as the
mass of the physical Higgs field $m_{h}=125~\mathrm{GeV}$ \cite{ATLAS, CMS}.
Also, the physical mutual coupling constant $\eta _{01}^{(4)}$ between
$S_{0} $ and $S_{1}$ is determined by the DM relic-density constraint \cite{Planck}, which
translates into the condition:
\begin{equation}
\left\langle v_{12}\sigma _{\mathrm{ann}}\right\rangle \simeq 1.7\times
10^{-9}\,\mathrm{GeV}^{-2},  \label{v12sigma_annihi}
\end{equation}
where $\left\langle v_{12}\sigma _{\mathrm{ann}}\right\rangle $ is the
thermally averaged annihilation cross-section of a pair of two DM particles
times their relative speed in the center-of-mass reference frame. This
constraint is imposed throughout this work, together with the perturbativity
restriction $0\leq \eta _{01}^{(4)}\leq \sqrt{4\pi }$ on its solution. The
remaining free parameters of the model are the physical mutual coupling
constant $\lambda _{0}^{(4)}$ between $h$ and $S_{0}$, the mixing angle 
$\theta $ between $h$ and $S_{1}$, the DM mass $m_{0}$, the mass $m_{1}$ of
the auxiliary physical field $S_{1}$, and the DM self-coupling constant
$\eta _{0}$. This latter has so far been decoupled from the other coupling
constants \cite{abada-ghaffor-nasri,abada-nasri-1}, but not anymore in view
of (\ref{beta-functions-scalars}) now that the running is the focus.
However, its initial value $\eta _{0}\left( \Lambda _{0}\right) $ is
arbitrary and its $\beta $-function is always positive. This means $\eta
_{0}\left( \Lambda \right) $ will only increase as $\Lambda $ increases,
quickly if starting from a rather large initial value, slowly if not.
Therefore, without loosing generality in the subsequent discussion, we fix
$\eta _{0}\left( \Lambda _{0}\right) =1$. Hence, here too we still
effectively have four free parameters: $\lambda _{0}^{(4)}$, $\theta $, 
$m_{0}$, and $m_{1}$. The initial conditions for the coupling constants in
(\ref{U}) in terms of these physical free parameters are as follows:
\begin{eqnarray}
\eta _{1}\left( \Lambda _{0}\right) &=&\frac{3}{2v_{1}^{2}}\left[
m_{1}^{2}+m_{h}^{2}+\left\vert m_{1}^{2}-m_{h}^{2}\right\vert \left( \cos
\left( 2\theta \right) +\frac{v}{2v_{1}}\sin \left( 2\theta \right) \right) \right] ;  \notag \\
\lambda \left( \Lambda _{0}\right) &=&\frac{3}{2v^{2}}\left[
m_{1}^{2}+m_{h}^{2}-\left\vert m_{1}^{2}-m_{h}^{2}\right\vert \left( \cos
\left( 2\theta \right) -\frac{v_{1}}{2v}\sin \left( 2\theta \right) \right) \right] ;  \notag \\
\lambda _{1}\left( \Lambda _{0}\right) &=&\frac{\sin \left( 2\theta \right) 
}{2vv_{1}}\left\vert m_{1}^{2}-m_{h}^{2}\right\vert ;  \notag \\
\eta _{01}\left( \Lambda _{0}\right) &=&\frac{1}{\cos \left( 2\theta \right) 
}\left[ \eta _{01}^{(4)}\cos ^{2}\theta -\lambda _{0}^{(4)}\sin ^{2}\theta \right] ;  \notag \\
\lambda _{0}\left( \Lambda _{0}\right) &=&\frac{1}{\cos \left( 2\theta
\right) }\left[ \lambda _{0}^{(4)}\cos ^{2}\theta -\eta _{01}^{(4)}\sin
^{2}\theta \right] .  \label{couplings-vs-parameters}
\end{eqnarray}
Note that normally, as we go down the mass scale, we should seam quantities
in steps: at $v$, $v_{1}$, and $\Lambda _{0}$. However, the corrections to
(\ref{couplings-vs-parameters}) are of one-loop order times $\ln
\frac{v}{v_{1}}$ or $\ln \frac{v_{1}}{\Lambda _{0}}$, small enough for our present
purposes to neglect.

\section{Scalars Only}

To see the effects of the scalar couplings only and how up in the mass scale
the model can go, we switch off the non-Higgs SM couplings in
(\ref{beta-functions-scalars}). The perturbativity constraint we impose on all
dimensionless scalar coupling constants is $G\left( \Lambda \right) \leq 
\sqrt{4\pi }$. Vacuum stability means that $G\left( \Lambda \right) \geq 0$
for the self-coupling constants $\eta _{0},\lambda $, and $\eta _{1}$, and
the conditions:
\begin{equation}
-\frac{1}{6}\sqrt{\eta _{0}\lambda }\leq \text{$\lambda _{0}$}\leq
\sqrt{4\pi };\qquad -\frac{1}{6}\sqrt{\eta _{0}\text{$\eta _{1}$}}\leq \text{$\eta
_{01}$}\leq \sqrt{4\pi };\qquad -\frac{1}{6}\sqrt{\text{$\eta _{1}$}\lambda }
\leq \text{$\lambda _{1}$}\leq \sqrt{4\pi }  \label{stability-conditions}
\end{equation}
for the mutual couplings $\lambda _{0},\eta _{01}$, and $\lambda _{1}$.
Also, as a start, we let the masses $m_{0}$ and $m_{1}$ vary in the interval 
$1\mathrm{GeV}-200\mathrm{GeV}$.

\begin{figure}[htb]
\centering
\includegraphics[width=6in,height=4in]{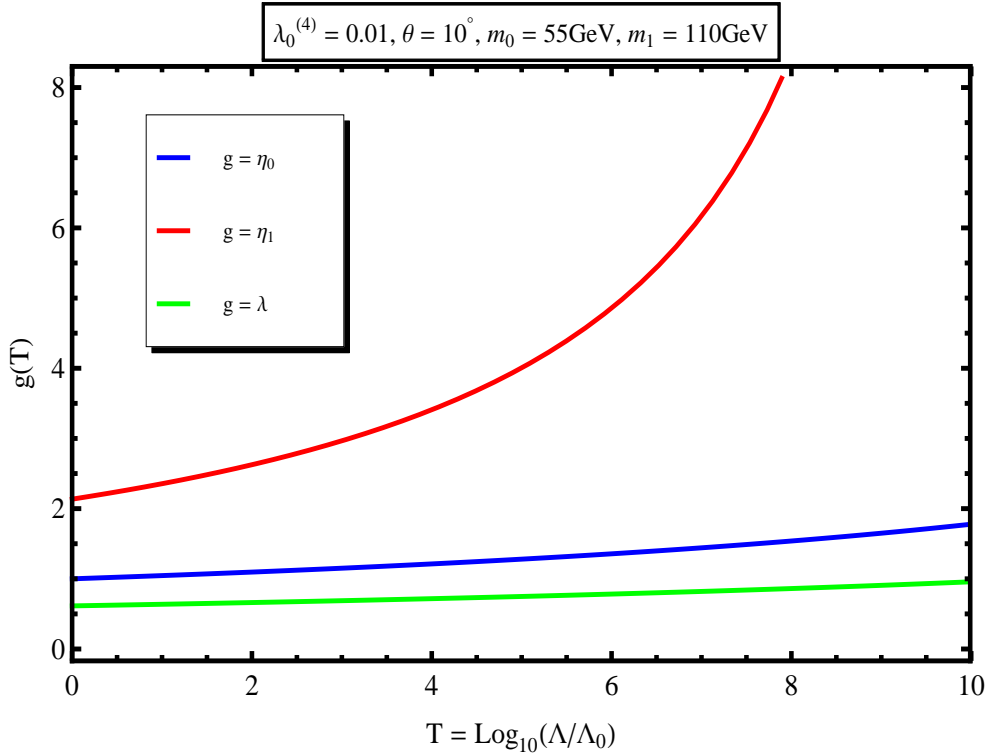}
\caption{The running of the self-couplings (scalars only). The self-coupling $\eta _{1}$ of the
auxiliary field $\chi _{1}$ dominates over the Higgs self-coupling $\lambda $.}
\label{RGE-scal-self_lambda04-001_theta-10_m0-55_m1-110}
\end{figure}

\begin{figure}[htb]
\centering
\includegraphics[width=6in,height=4in]{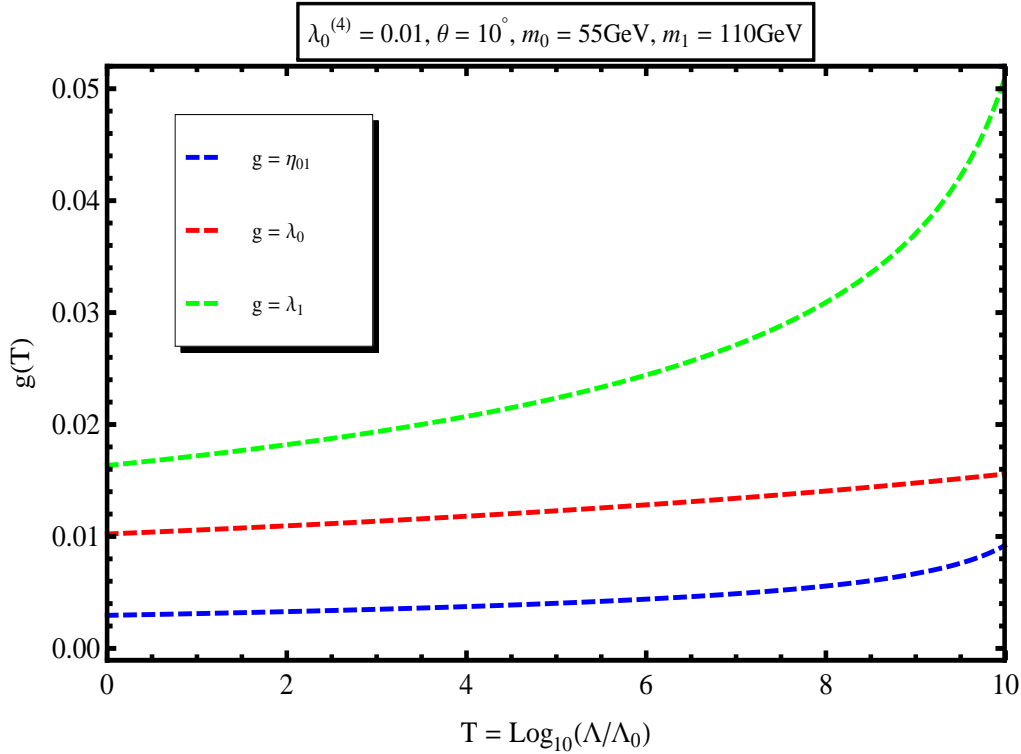}
\caption{Running of the mutual couplings (scalars only). For these values of the parameters, all three are
well below the self-couplings.}
\label{RGE-scal-mutual_lambda04-001_theta-10_m0-55_m1-110}
\end{figure}

Fig.~\ref{RGE-scal-self_lambda04-001_theta-10_m0-55_m1-110} displays a
typical running of the scalar self-coupling constants, from $\Lambda
_{0}=10^{2}\mathrm{GeV}$ up to $10^{12}\mathrm{GeV}$. As is expected for
scalars only, all coupling constants are increasing functions of the scale 
$\Lambda $, with different but increasing rates. Also, the larger value the
coupling starts from at $\Lambda _{0}$, the faster it will go up.
Fig.~\ref{RGE-scal-mutual_lambda04-001_theta-10_m0-55_m1-110} shows the running of
the mutual coupling constants for the same values of the parameters. For
these values, the mutual coupling constants start well below 1, and so run
low; they are very much dominated by the self-couplings. This situation will
stay for $\lambda _{0}$ and $\lambda _{1}$ in all regions, but not for $\eta
_{01}$.

The first coupling constant that leaves the perturbativity bound $\sqrt{4\pi 
}$ is $\eta _{1}$, the self-coupling constant of the auxiliary scalar field
$\chi _{1}$, at about $1260~\mathrm{TeV}$ for this set of values of the
parameters. This behavior is in fact typical. Indeed, $\eta _{1}$ starts
above 2 at $\Lambda _{0}$ in all the parameter space, much higher than all
the other coupling constants -- only $\eta _{01}$ can compete with it in
some regions. As it intervenes squared in its own $\beta $-function, it will
also move up quicker. More precisely, from (\ref{couplings-vs-parameters}),
we see that $\eta _{1}\left( \Lambda _{0}\right) $ depends on $m_{1}$ and
$\theta $ only. The effect of the mixing angle $\theta $ is small. As a
function of $m_{1}$, starting from about 2, $\eta _{1}\left( \Lambda
_{0}\right) $ decreases slightly until $m_{h}$ and then picks up. It will
pass the perturbativity bound $\sqrt{4\pi }$ at about $m_{1}\simeq
160\mathrm{GeV}$. This means that the region $m_{1}>160\mathrm{GeV}$ is
automatically excluded from the outset. In actual situations, given the
positive-slope RG running of $\eta _{1}\left( \Lambda \right) $ and even in
the case of the full RGE\ (see below), perturbativity puts a stricter upper
bound on $m_{1}$, irrespective of the other parameters of the model.

In Fig.~\ref{RGE-scal-self_lambda04-001_theta-10_m0-55_m1-110}, the Higgs
self-coupling $\lambda $ starts just above 0.6 and does not pick up much
when running. This behavior is typical too. Indeed, $\lambda \left( \Lambda
_{0}\right) $ is also a function of $m_{1}$ and $\theta $ only, see
(\ref{couplings-vs-parameters}). For a given $\theta $, it will increase as a
function of $m_{1}$ to reach $3\left( m_{h}/v\right) ^{2}$ at $m_{1}=m_{h}$,
equal here to $0.77$ for $m_{h}=125\mathrm{GeV}$. Then it continues to
increase, but with a smaller slope. The mixing angle $\theta $ enhances the
behavior of $\lambda \left( \Lambda _{0}\right) $ as a function of $m_{1}$,
but for, say\footnote{We are implicitly confining the mixing angle $\theta $ to small values, a
situation inferred from our work \cite{abada-nasri-1} on the
phenomenological implications of the model. This is discussed later in
section V.} $\theta =15^{\mathrm{o}}$, $\lambda \left( \Lambda _{0}\right) $
will be less than $0.85$ at $m_{1}=160~\mathrm{GeV}$. This situation implies
that when running as a function of the scale $\Lambda $, the Higgs
self-coupling $\lambda $ will increase, but will hardly reach 1 before, say 
$\eta _{1}$, leaves the perturbativity bound.

\begin{figure}[htb]
\centering
\includegraphics[width=6in,height=4in]{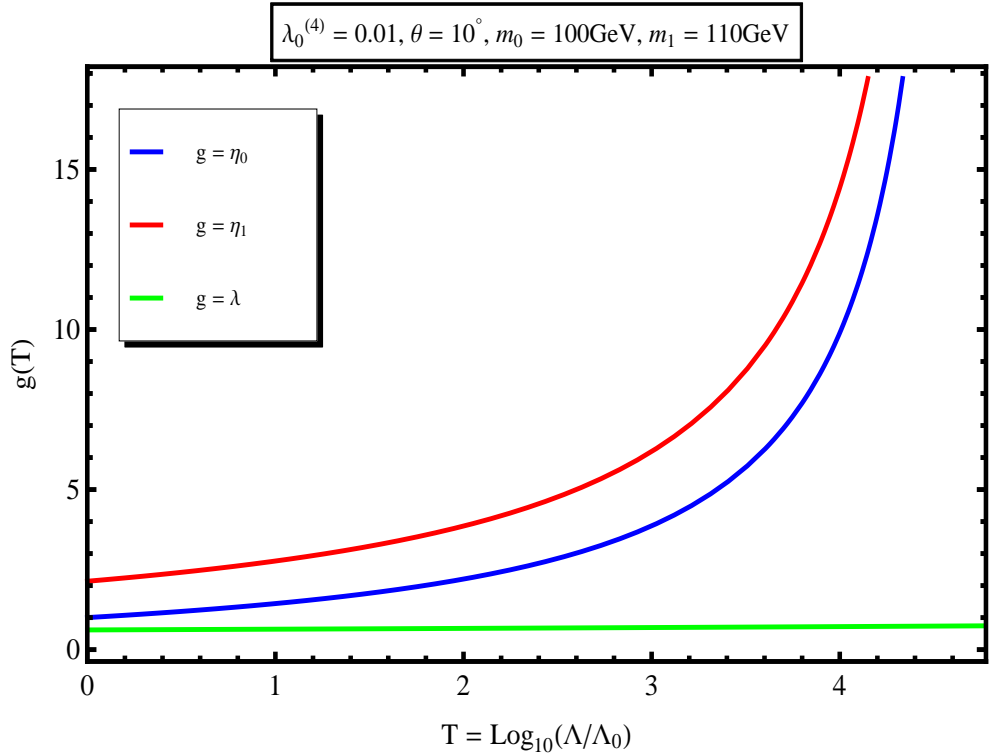}
\caption{Running of the self-couplings (scalars only) for a larger DM mass. $\eta _{1}$ is still dominant,
even if $\eta _{0}$ picks up faster behind it.}
\label{RGE-scal-self_lambda04-001_theta-10_m0-100_m1-110}
\end{figure}

\begin{figure}[htb]
\centering
\includegraphics[width=6in,height=4in]{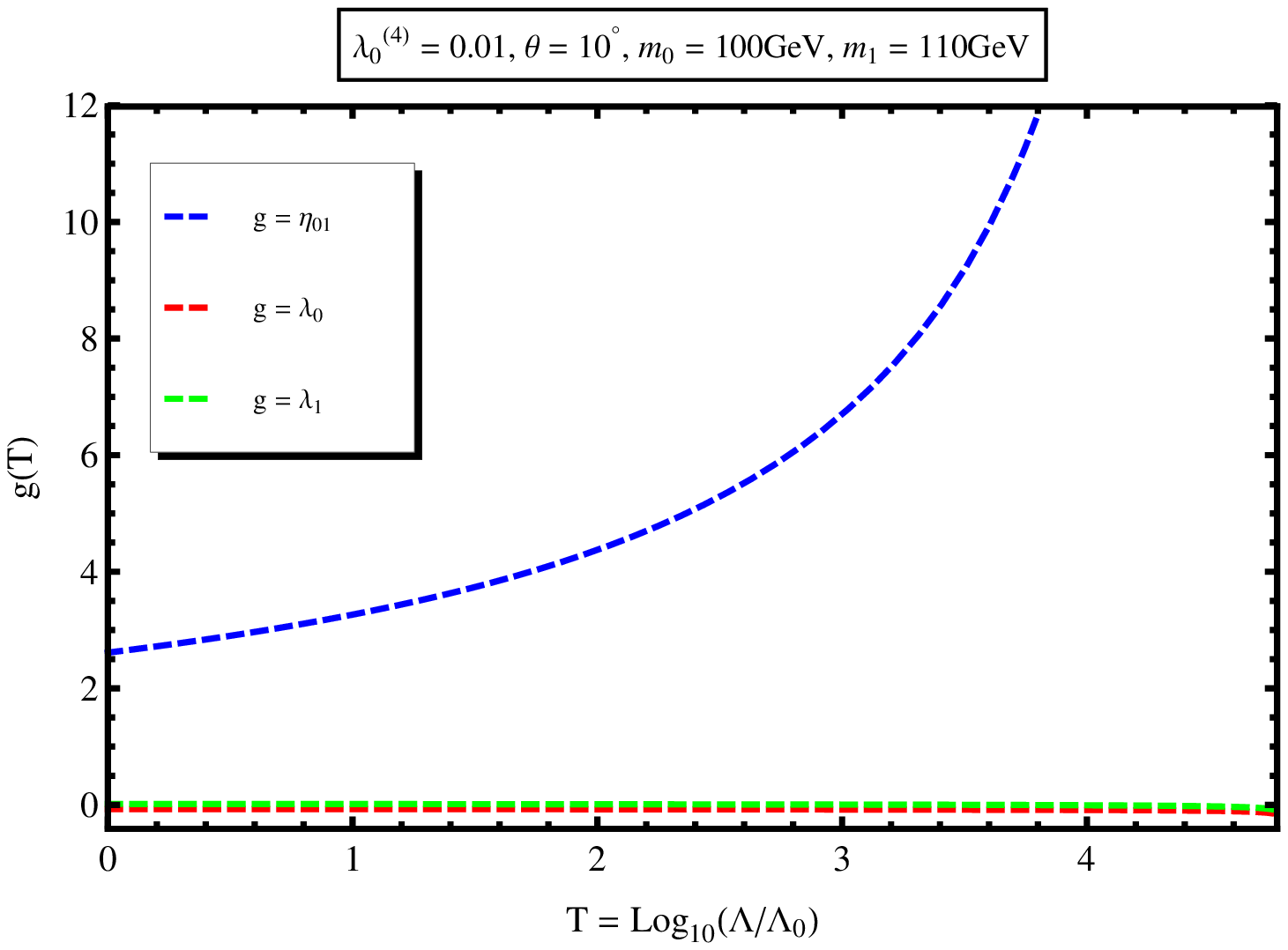}
\caption{Running of the mutual couplings (scalars only) for a larger DM mass. $\eta _{01}$ starts
above 1, much higher than the two others, even higher than the self-coupling $\eta _{1}$.}
\label{RGE-scal-mutual_lambda04-001_theta-10_m0-100_m1-110}
\end{figure}

Increasing the dark-matter mass does affect the running of the couplings.
Figs.~\ref{RGE-scal-self_lambda04-001_theta-10_m0-100_m1-110}
(self-couplings) and \ref{RGE-scal-mutual_lambda04-001_theta-10_m0-100_m1-110} (mutual couplings)
display such effects for $m_{0}=100\mathrm{GeV}$. Among the self-couplings,
$\eta _{1}$ is still dominant, but tailed more closely by $\eta _{0}$ this
time. For both, the positive acceleration is accentuated, something that
makes $\eta _{1}$ leave the perturbativity region much earlier, at about 
$6.3\mathrm{TeV}$. By contrast, the running of the Higgs self-coupling $\lambda $
stays flat.

The major effect of increasing $m_{0}$ is on the mutual coupling $\eta _{01}$, 
between the DM field $S_{0}$ and the auxiliary field $\chi _{1}$. Indeed,
in Fig.~\ref{RGE-scal-mutual_lambda04-001_theta-10_m0-55_m1-110} where 
$m_{0}=55\mathrm{GeV}$, $\eta _{01}$ started and ran small like the other two
mutual couplings. Here, whereas $\lambda _{0}$ and $\lambda _{1}$ (both
Higgs related) stay close to zero, $\eta _{01}$ starts above 2.5 and runs up
fast. In fact, for these values of the parameters, it leaves the
perturbativity region earlier than $\eta _{1}$, at about $2.5\mathrm{TeV}$.

\begin{figure}[htb]
\centering
\includegraphics[width=6in,height=4in]{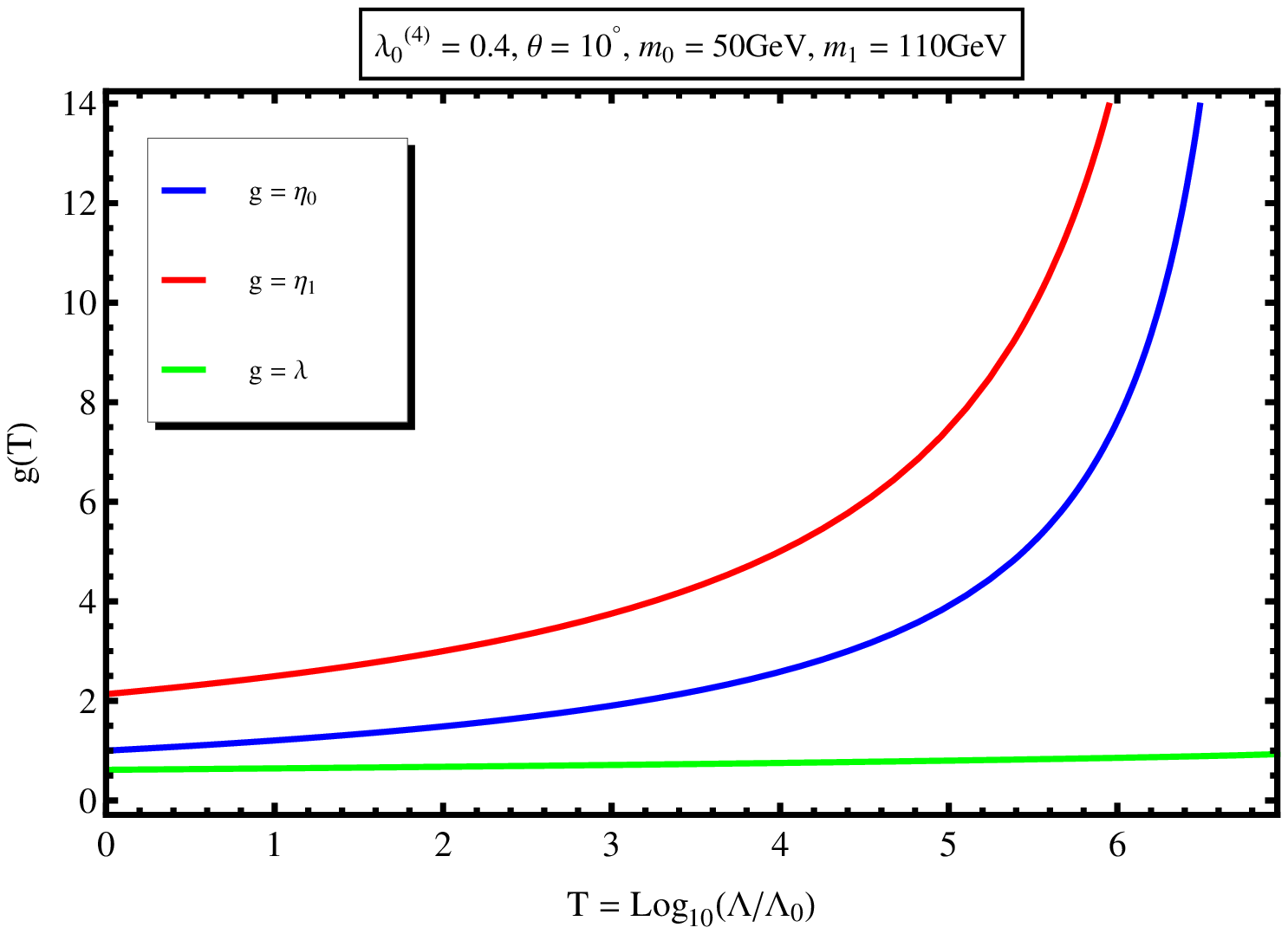}
\caption{Running of the self-couplings (scalars only) with a larger physical Higgs self-coupling
$\lambda _{0}^{(4)}$. The self-coupling $\eta _{1}$ is still dominant.}
\label{RGE-scal-self_lambda04-04_theta-10_m0-50_m1-110}
\end{figure}

Increasing the auxiliary-field mass $m_{1}$ has a similar effect: it
enhances the positive acceleration of the self-couplings $\eta _{1}$ and
$\eta _{0}$ while leaving $\lambda $ flat, and boosts up the mutual coupling
$\eta _{01}$ away from $\lambda _{0}$ and $\lambda _{1}$, which both remain
not far from zero. It also makes $\eta _{1}$ and $\eta _{01}$ leave the
perturbativity region earlier, without $\eta _{01}$ necessarily taking over
from $\eta _{1}$.

\begin{figure}[htb]
\centering
\includegraphics[width=6in,height=4in]{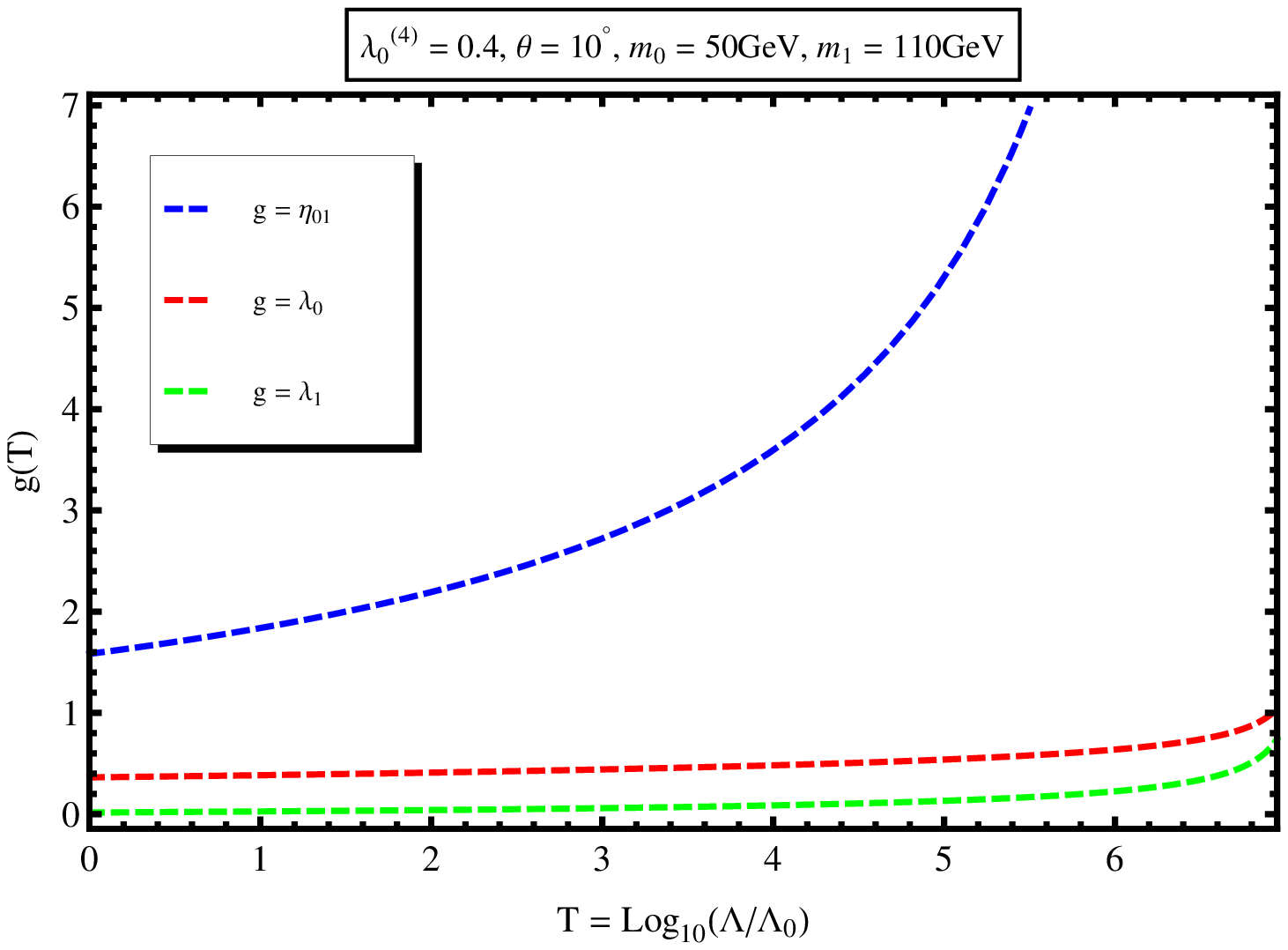}
\caption{Running of mutual couplings (scalars only). Larger $\lambda _{0}^{(4)}$ helps
$\eta _{01} $ rise well above $\lambda _{0}$ and $\lambda _{1}$,
but not enough to win over the self-coupling $\eta _{1}$.}
\label{RGE-scal-mutual_lambda04-04_theta-10_m0-50_m1-110}
\end{figure}

Increasing $\lambda _{0}^{(4)}$ has also an effect. 
Figs.~\ref{RGE-scal-self_lambda04-04_theta-10_m0-50_m1-110} (self) and
\ref{RGE-scal-mutual_lambda04-04_theta-10_m0-50_m1-110} (mutual) show the
running for $\lambda _{0}^{(4)}=0.4$. The self-coupling $\eta _{1}$
dominates and leaves the perturbativity region at about $251\mathrm{TeV}$.
The mutual coupling $\eta _{01}$ is raised above 1 at $\Lambda _{0}$\ and so
can run high, while $\lambda _{0}$ and $\lambda _{1}$ stay here too just
above zero. It leaves the perturbativity region at about $794\mathrm{GeV}$,
well behind $\eta _{1}$. Higher values of $\lambda _{0}^{(4)}$ are more
difficult to achieve as the relic-density constraint (\ref{v12sigma_annihi})
may not be satisfied \cite{abada-ghaffor-nasri}.

Finally, changing the mixing angle $\theta $ has little effect on the
self-coupling constants. It helps the mutual coupling constants $\eta _{01}$
and $\lambda _{1}$ start higher, but not by much: they stay with $\lambda
_{0}$ well below one.

\section{The full RGE}

\begin{figure}[htb]
\centering
\includegraphics[width=6in,height=4in]{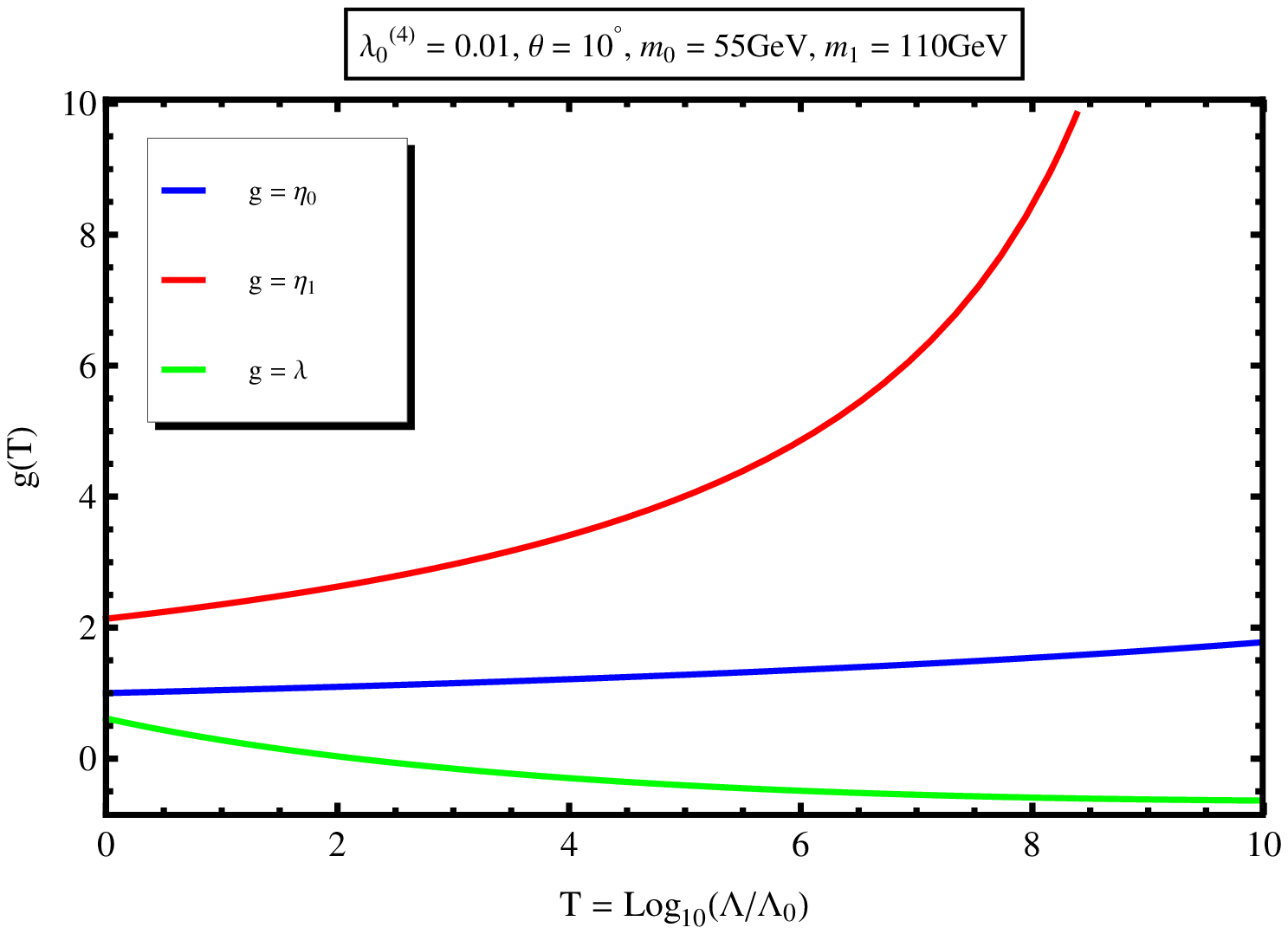}
\caption{Running of the self-couplings (full RGE). $\eta _{1}$ controls perturbativity and the Higgs
coupling $\lambda $ becomes negative quickly.}
\label{RGE-self_lambda04-001_theta-10_m0-55_m1-110}
\end{figure}

\begin{figure}[htb]
\centering
\includegraphics[width=6in,height=4in]{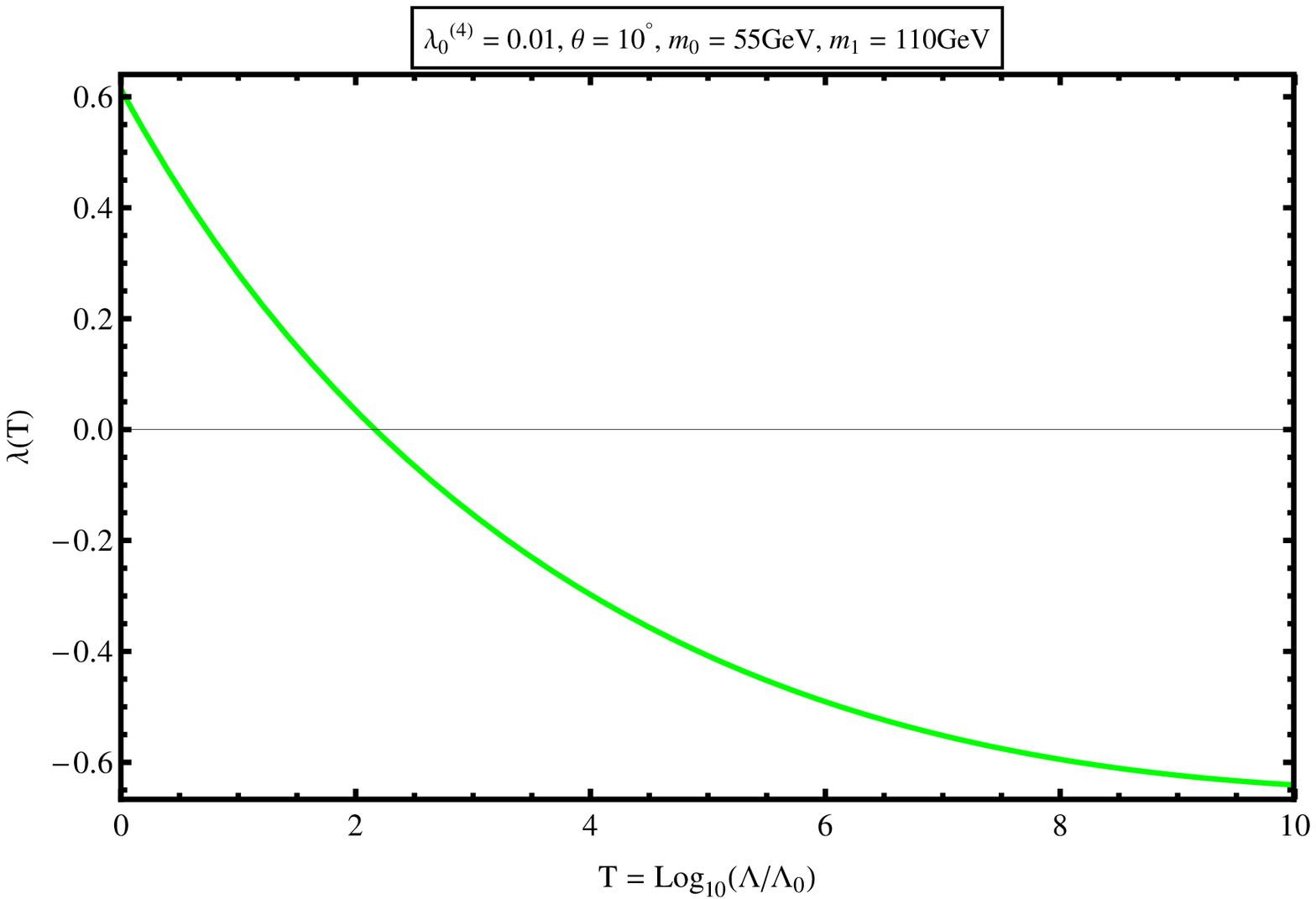}
\caption{The running of the Higgs self-coupling $\lambda $ (full RGE). It gets negative at about $15 
\mathrm{TeV}$ for this set of parameters' values.}
\label{RGE-lambda_lambda04-001_theta-10_m0-55_m1-110}
\end{figure}

In the previous situation, `scalars only', all running coupling constants
were positive, and so there were no issues related to vacuum stability. We
now reintroduce the other SM\ particles and see their effects.
Fig.~\ref{RGE-self_lambda04-001_theta-10_m0-55_m1-110} displays the behavior of the
self-couplings under the full RGE for the same values of the parameters as
in Fig.~\ref{RGE-scal-self_lambda04-001_theta-10_m0-55_m1-110} (scalars
only). The dramatic effect is on the Higgs self-coupling constant $\lambda $
which quickly gets into negative territory, at about $15\mathrm{TeV}$, thus
rendering the theory unstable beyond this mass scale. This is better
displayed in Fig.~\ref{RGE-lambda_lambda04-001_theta-10_m0-55_m1-110} where
the RG behavior of $\lambda $ is shown by itself. Such a negative slope for
$\lambda $ is expected, given the negative contributions to $\beta _{\lambda
} $ in (\ref{beta-functions-scalars}). Here too $\eta _{1}$ is dominant\
over the other couplings and still controls perturbativity, leaving its
region much later, at about $1600~\mathrm{TeV}$, farther away from the
situation `scalars only'. This looks to be a somewhat general trend: the
non-Higgs SM particles seem to flatten the runnings of the scalar couplings.

\begin{figure}[htb]
\centering
\includegraphics[width=6in,height=4in]{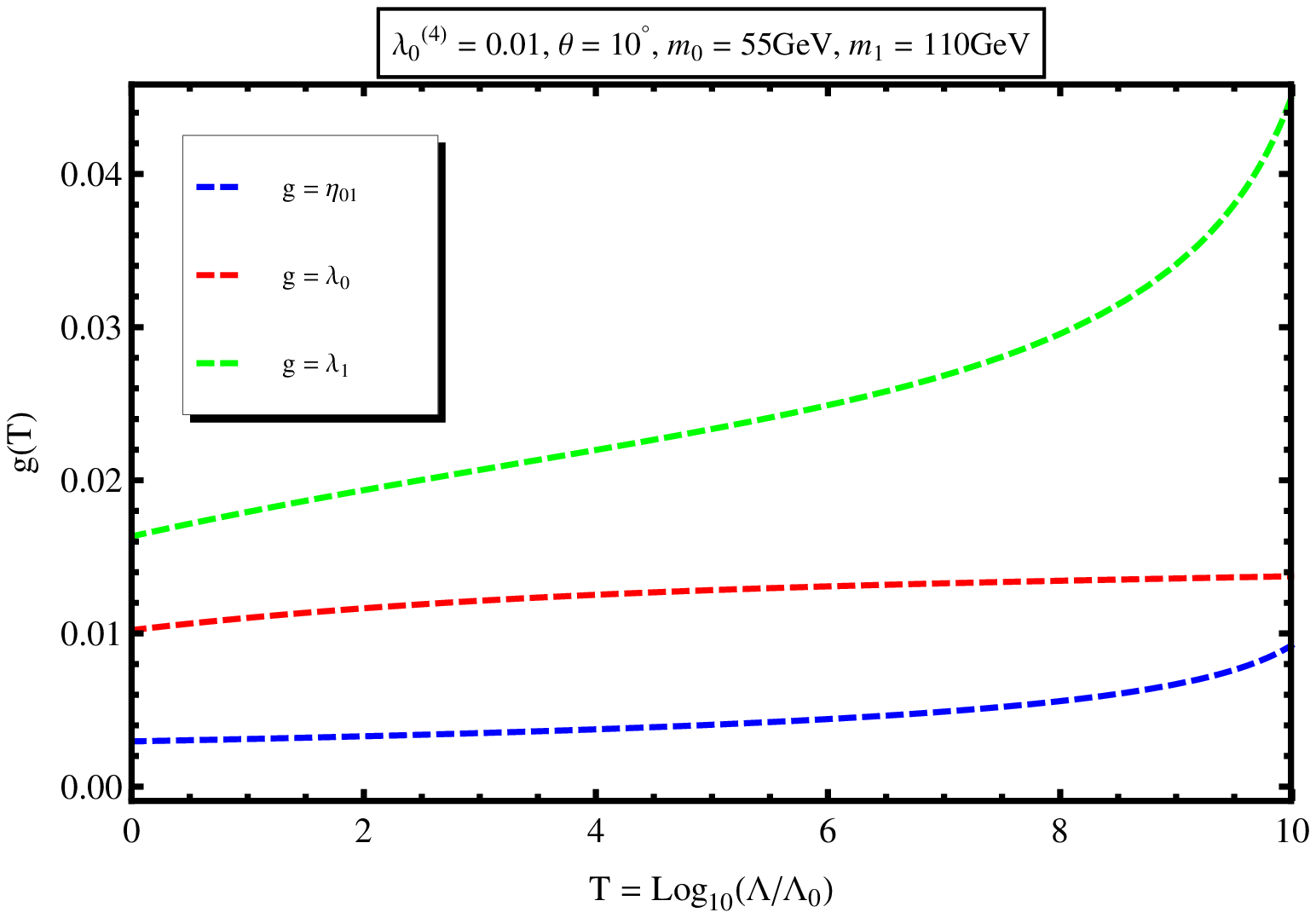}
\caption{Running of the mutual couplings (full RGE). The inclusion of the other SM particles flattens the
runnings.}
\label{RGE-mutual_lambda04-001_theta-10_m0-55_m1-110}
\end{figure}

\begin{figure}[htb]
\centering
\includegraphics[width=6in,height=4in]{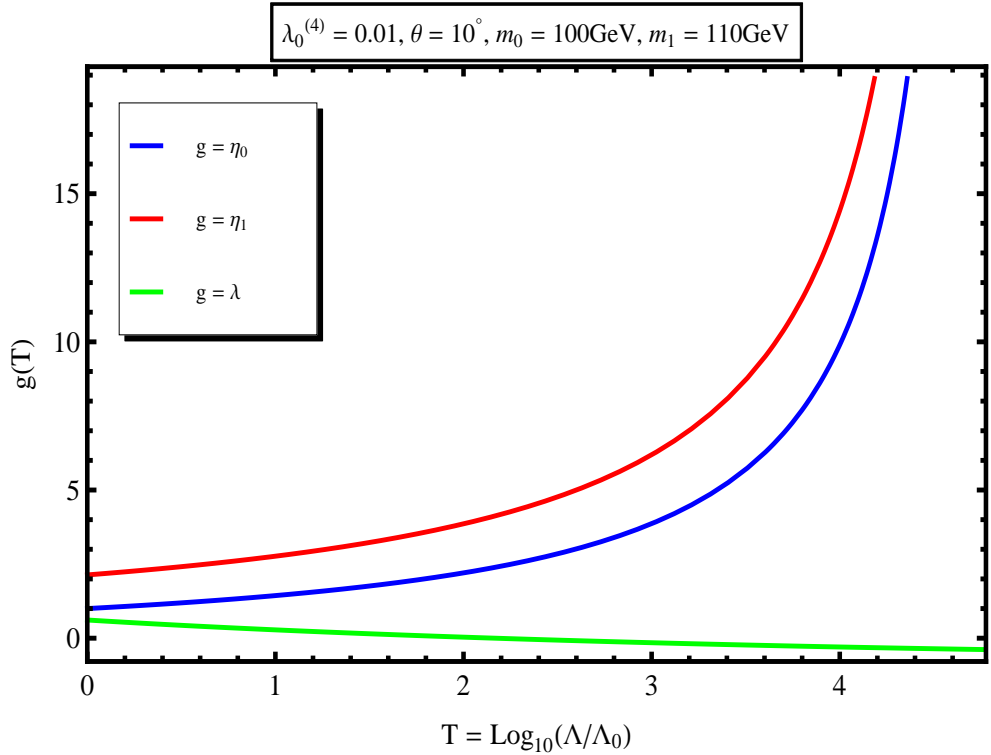}
\caption{Running of the self-couplings (full RGE) with $m_{0}$ larger. $\eta _{1}$ is more closely tailed
by $\eta _{0}$ and $\lambda $ decreases and turns negative
at about 10TeV.}
\label{RGE-self_lambda04-001_theta-10_m0-100_m1-110}
\end{figure}

\begin{figure}[htb]
\centering
\includegraphics[width=6in,height=4in]{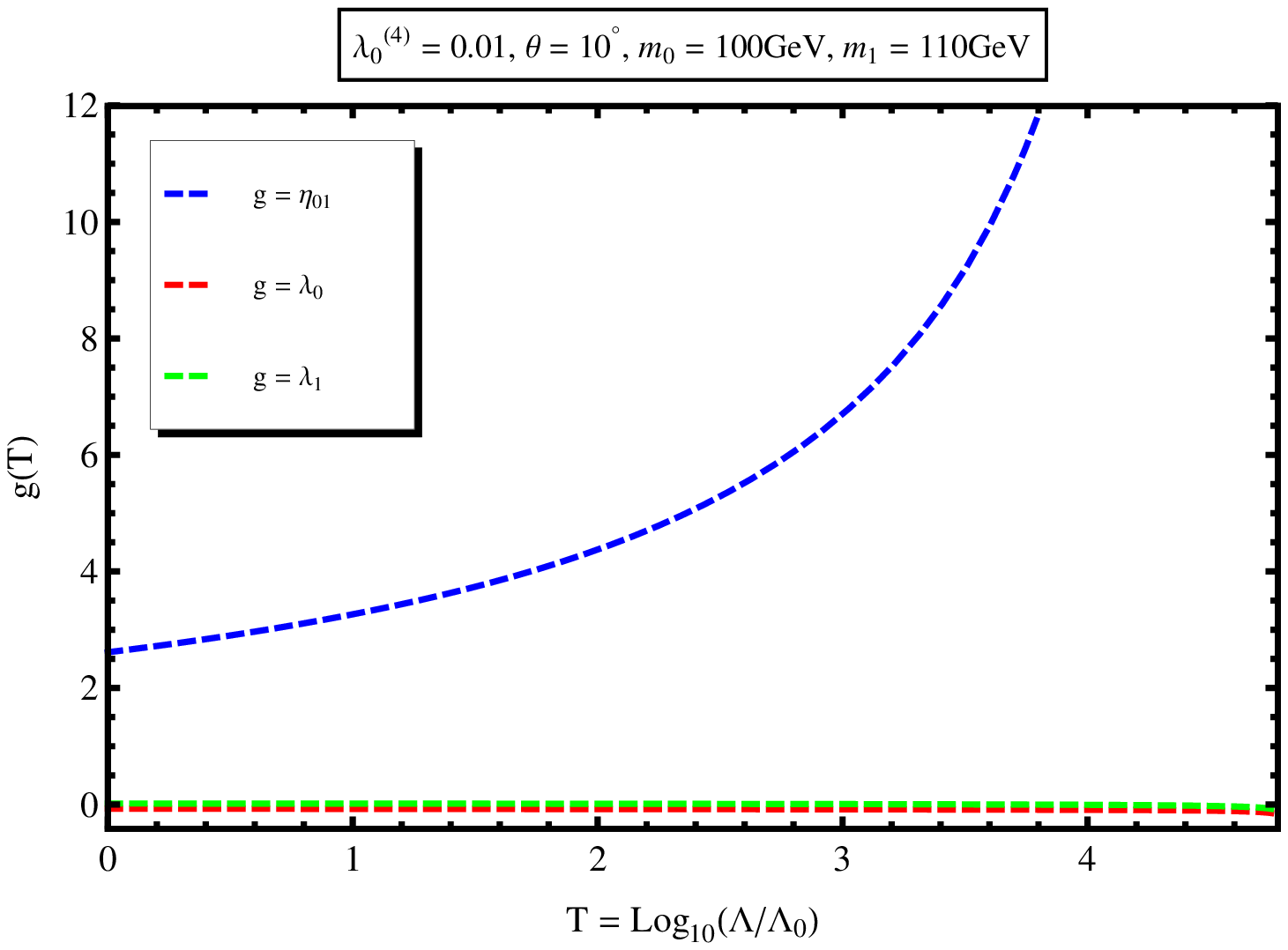}
\caption{Running of the mutual couplings (full RGE) with $m_{0}$ larger. $\eta _{01}$ starts well
above $\lambda _{0}$ and $\lambda _{1}$ and leaves the
perturbativity region before the self-coupling $\eta _{1}$.}
\label{RGE-mutual_lambda04-001_theta-10_m0-100_m1-110}
\end{figure}

The runnings of the mutual coupling constants for the same set of
parameters' values is displayed in Fig.~\ref{RGE-mutual_lambda04-001_theta-10_m0-55_m1-110}.
They also get flattened by the other SM particles, but they stay positive. Here too they dwell well
below the self-couplings, as in the `scalars only' case. In fact, many of
the effects on the running coupling constants coming from varying the
parameters are similar to those of the previous situation since the SM
particles do not intervene in the initial values of the couplings (self and
mutual) at $\Lambda _{0}$. This means that increasing $m_{0}$ and $m_{1}$
will raise the mutual coupling $\eta _{01}$ and not the two others, higher
than $\eta _{1}$ in some regions. For example,
Fig.~\ref{RGE-self_lambda04-001_theta-10_m0-100_m1-110} shows the running of the
self-couplings when $m_{0}=100\mathrm{GeV}$. Both $\eta _{1}$ and $\eta _{0}$
run faster but $\lambda $ is little affected.
Fig.~\ref{RGE-mutual_lambda04-001_theta-10_m0-100_m1-110} shows the running of the
mutual couplings from the full RGE also at $m_{0}=100\mathrm{GeV}$. As in
the case `scalars only', larger $m_{0}$ boosts up $\eta _{01}\left( \Lambda
_{0}\right) $, much higher than $\lambda _{0}$ and $\lambda _{1}$, at about
2.2 here, which makes it run quickly high, leaving the perturbativity region
before $\eta _{1}$, as in the case `scalars only'.

Raising $\lambda _{0}^{(4)}$ will also make the self-couplings $\eta _{1}$
and $\eta _{0}$ run faster while affecting very little $\lambda $. It will
also make the mutual coupling $\eta _{01}$ starts higher, and so demarked
from $\lambda _{0}$ and $\lambda _{1}$. By contrast, the effect of $\theta $
is not very dramatic: the self-couplings are not much affected and the
mutuals only evolve differently, without any particular boosting of $\eta
_{01}$.

\section{Regions of Viability}

The foregoing discussion showed us how the scalar parameters of the
two-singlet model behave as a function of the mass scale $\Lambda $. From
the situation `scalars only' we understood that the two couplings that
control perturbativity are $\eta _{1}$ and $\eta _{01}$. The full RGE
brought in stability: the change of sign of $\lambda $ is the vacuum
stability criterion to use. Equipped with these indicators, we can try to
investigate in a more systematic way the viability regions of the model,
regions in the space of parameters in which the model is predictive.
Remember that this model has four parameters: the dark-matter mass $m_{0}$,
the physical auxiliary field mass $m_{1}$, the physical Higgs self-coupling 
$\lambda _{0}^{(4)}$, and the mixing angle $\theta $ between the physical
Higgs and the auxiliary field. The way we proceed is to vary $\lambda
_{0}^{(4)}$ and $\theta $ and try to find the regions of viability of the
model in the $\left( m_{0},m_{1}\right) $-plane.

We have by now a number of tools at our disposal. First the DM relic-density
constraint (\ref{v12sigma_annihi}), which has been applied throughout and
will continue so. We have the RGE analysis of this work. We will require
both $\eta _{1}\left( \Lambda \right) $ and $\eta _{01}\left( \Lambda
\right) $ to be smaller than $\sqrt{4\pi }$, and $\lambda \left( \Lambda
\right) $ to be positive.

\begin{figure}[htb]
\centering
\includegraphics[width=5.5in,height=5in]{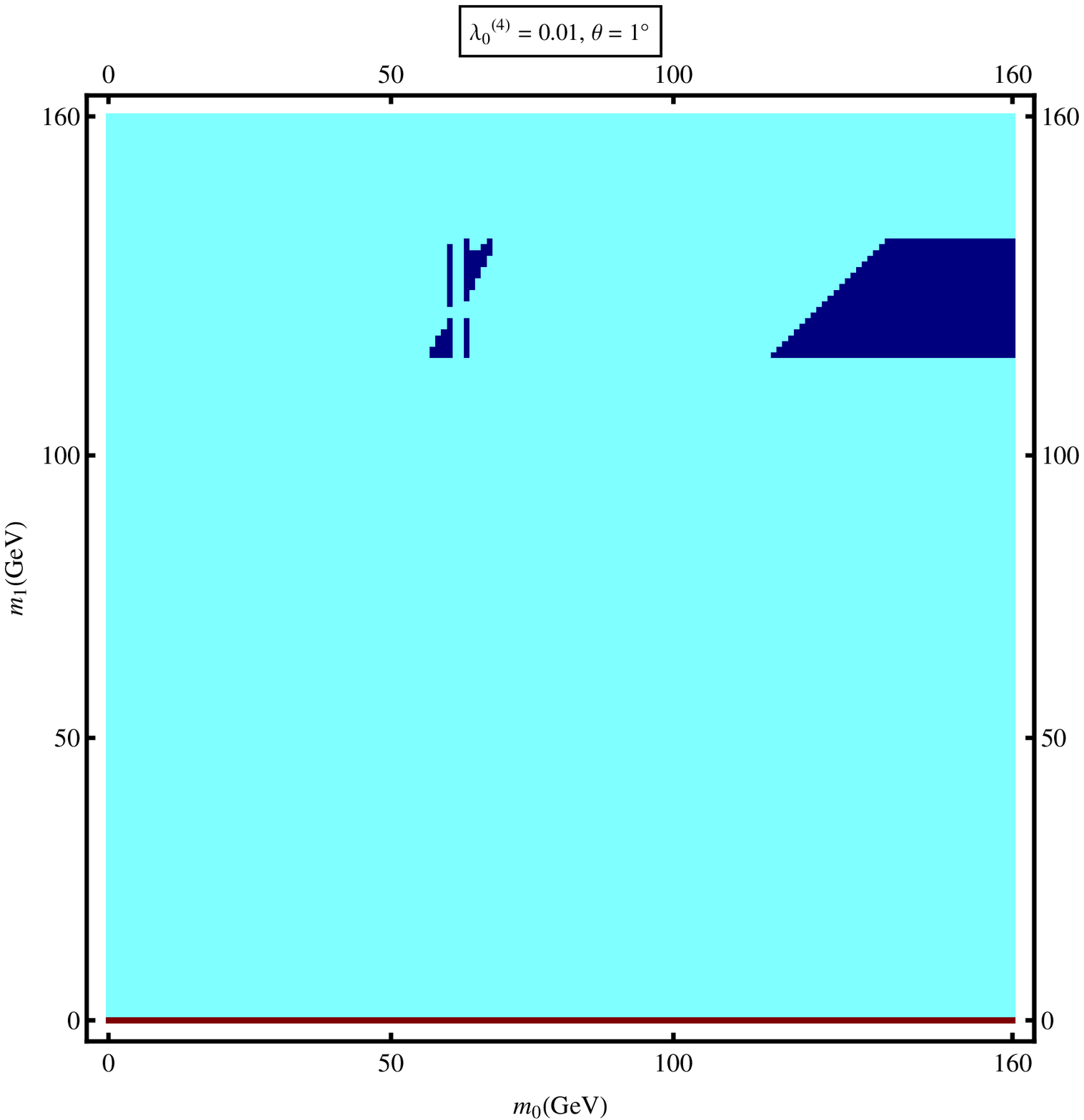}
\caption{Regions of viability of the two-singlet model (in blue). Physical Higgs self-coupling
$\lambda _{0}^{(4)}$ and mixing angle $\protect\theta $ very small.}
\label{RGE-dens10_lambda04-001_theta-1_tm-6_v2}
\end{figure}

There is one important issue to address though before we proceed, and that
is how far we want the model to be perturbatively predictive and stable. The
maximum value $\Lambda _{m}$ for the mass scale $\Lambda $ should not be
very high for two reasons. One, more conceptual, is that we want to
recognize and allow the model to be intermediary between the Standard Model
and some possible higher structure. The second reason, more practical, is
that a too-high $\Lambda _{m}$ is too restrictive for the parameters
themselves. For example, for the parameters we used in the previous
sections, in particular $m_{0}=55\mathrm{GeV}$ and $m_{1}=110~\mathrm{GeV}$,
we have seen that $\lambda $ gets negative already for $\Lambda \simeq 15
\mathrm{TeV}$ whereas $\eta _{1}$ leaves the perturbativity region much
later, for $\Lambda \simeq 1600\mathrm{TeV}$. The situation can be
reversed. For example, for $m_{0}=67\mathrm{GeV}$, $m_{1}=135\mathrm{GeV}$,
and $\theta =15^{\mathrm{o}}$, $\lambda $ can live positive until about 
$400\mathrm{TeV}$ whereas $\eta _{1}$ leaves perturbativity at about
$50\mathrm{TeV}$. In this section, we set $\Lambda _{m}\simeq 40\mathrm{TeV}$.

\begin{figure}[htb]
\centering
\includegraphics[width=5.5in,height=5in]{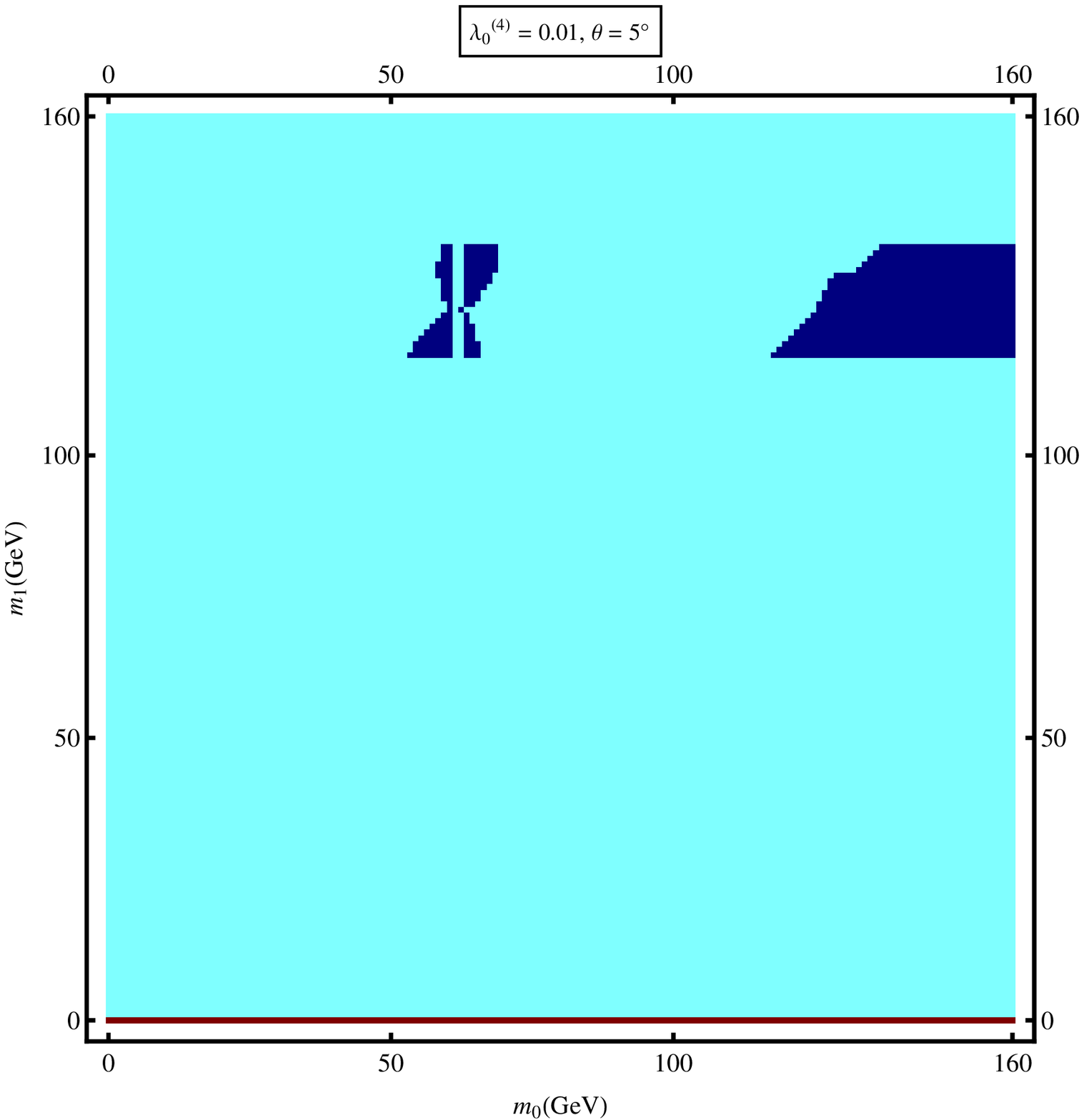}
\caption{Regions of viability (blue) of the model. $\lambda _{0}^{(4)}$ still very small, but
$\theta $ larger. The region is richer, but not relocated.}
\label{RGE-dens10_lambda04-001_theta-5_tm-6_v2}
\end{figure}

As a third viability tool, we want the model to comply with the measured
direct-detection upper bounds. In our model, the total cross section for
non-relativistic elastic scattering of a dark matter WIMP off a nucleon
target is given by the relation \cite{abada-ghaffor-nasri}: 
\begin{equation}
\sigma _{\det }=\frac{m_{N}^{2}\left( m_{N}-\frac{7}{9}m_{B}\right) ^{2}}
{4\pi \left( m_{N}+m_{0}\right) ^{2}v^{2}}\left[ \frac{\lambda _{0}^{\left(
3\right) }\cos {\theta }}{m_{h}^{2}}-\frac{\eta _{01}^{\left( 3\right) }\sin 
{\theta }}{m_{1}^{2}}\right] ^{2}.  \label{sigma_det}
\end{equation}
In this relation, $m_{N}$ is the nucleon mass and $m_{B}$ the baryon mass in
the chiral limit. The quantities $\lambda _{0}^{\left( 3\right) }$ and $\eta
_{01}^{\left( 3\right) }$ are coupling constants of cubic terms in the
theory after spontaneous breaking of the two symmetries \cite{abada-ghaffor-nasri}:
\begin{equation}
\lambda _{0}^{\left( 3\right) }=\lambda _{0}\cos \theta +\eta _{01}v_{1}\sin
\theta ;\quad \eta _{01}^{\left( 3\right) }=\eta _{01}v_{1}\cos \theta
-\lambda _{0}v\sin \theta .  \label{cubic-couplings}
\end{equation}
The condition we impose is that $\sigma _{\det }$ be within the XENON 100
upper-bounds \cite{XENON100}.

In work \cite{abada-nasri-1}, we studied phenomenological implications of
the model and constraints on it, using rare meson decays and Higgs
production. A number of inferences were deduced, but we will prefer to
retain only two. One is that the mixing angle $\theta $ is to be chosen
small. This is emphasized in view of the mounting evidence of a SM Higgs
particle found by ATLAS\ and CMS at the LHC \cite{ATLAS, CMS}. The other is
that the physical self-coupling $\lambda _{0}^{(4)}$ is to be small too.
This was already observed in \cite{abada-ghaffor-nasri}, where the
relic-density constraint has the tendency of `shutting down' high values of
$\lambda _{0}^{(4)}$. At the end of the next section, we will comment on possible
larger values for $\lambda _{0}^{(4)}$.

\begin{figure}[htb]
\centering
\includegraphics[width=5.5in,height=5in]{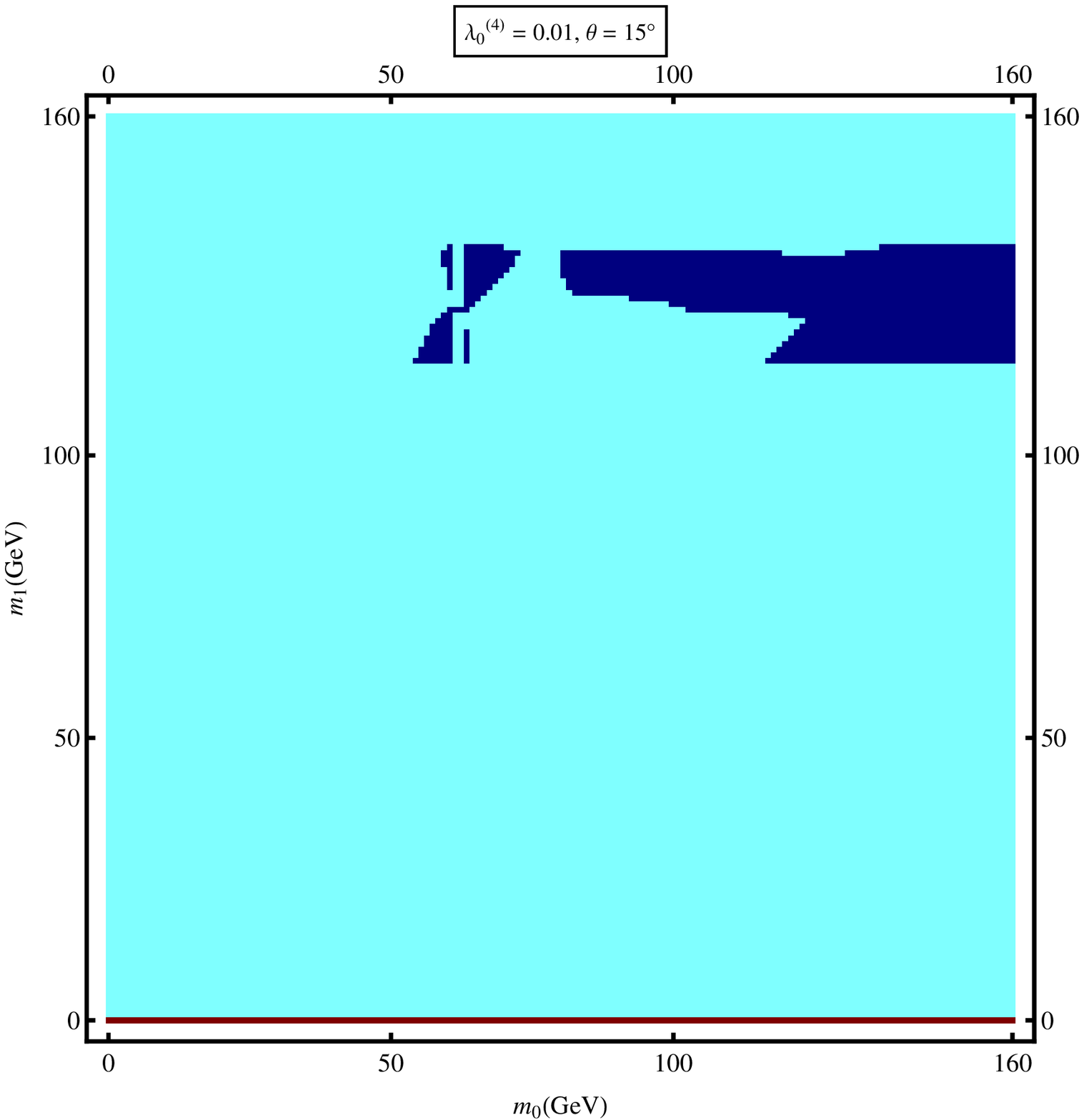}
\caption{The region of viability (blue) is even richer for larger mixing angle $\theta $.}
\label{RGE-dens10_lambda04-001_theta-15_tm-6_v2}
\end{figure}

In this section, the display range of $m_{0}$ and $m_{1}$ is from 
$1\mathrm{GeV}$ to $160\mathrm{GeV}$. Indeed, there is no reliable data to discuss
regarding a dark-matter mass below the GeV, and in view of the behavior of
$\eta _{1}$ at $\Lambda _{0}$ as a function of $m_{1}$, taking this latter
beyond $160\mathrm{GeV}$ is outside the perturbativity region. In practice,
$m_{0}$ was taken up to $200\mathrm{GeV}$, with no additional features to
report.

Let us start with $\lambda _{0}^{(4)}$ and $\theta $ both very small.
Fig.~\ref{RGE-dens10_lambda04-001_theta-1_tm-6_v2} displays the regions (blue)
for which the model is viable up to $\Lambda _{m}\simeq 40\mathrm{TeV}$.
Here $\lambda _{0}^{(4)}=0.01$ and $\theta =1^{\mathrm{o}}$. We see that the
mass $m_{1}$ is confined to the interval $116\mathrm{GeV}-138\mathrm{GeV}$.
The dark-matter mass is confined mainly to the region above $118\mathrm{GeV}$,
the left boundary of which having a positive slope as $m_{1}$ increases.
The DM mass $m_{0}$ has also a small showing in the narrow interval
$57\mathrm{GeV}-68\mathrm{GeV}$.

The effect of increasing the mixing angle $\theta $ is to enrich the
existing regions without relocating them. This is displayed in
Figs.~\ref{RGE-dens10_lambda04-001_theta-5_tm-6_v2} and
\ref{RGE-dens10_lambda04-001_theta-15_tm-6_v2} for which $\theta $ is increased
to $5^{\mathrm{o}}$ and $15^{\mathrm{o}}$ respectively. We see that, as $\theta $
increases, the region between the narrow band and the larger one to
the right gets populated. This means more dark-matter masses above
$60\mathrm{GeV}$ are allowed, but $m_{1}$ stays in the same interval, roughly
$116\mathrm{GeV}-138\mathrm{GeV}$.

\begin{figure}[htb]
\centering
\includegraphics[width=5.5in,height=5in]{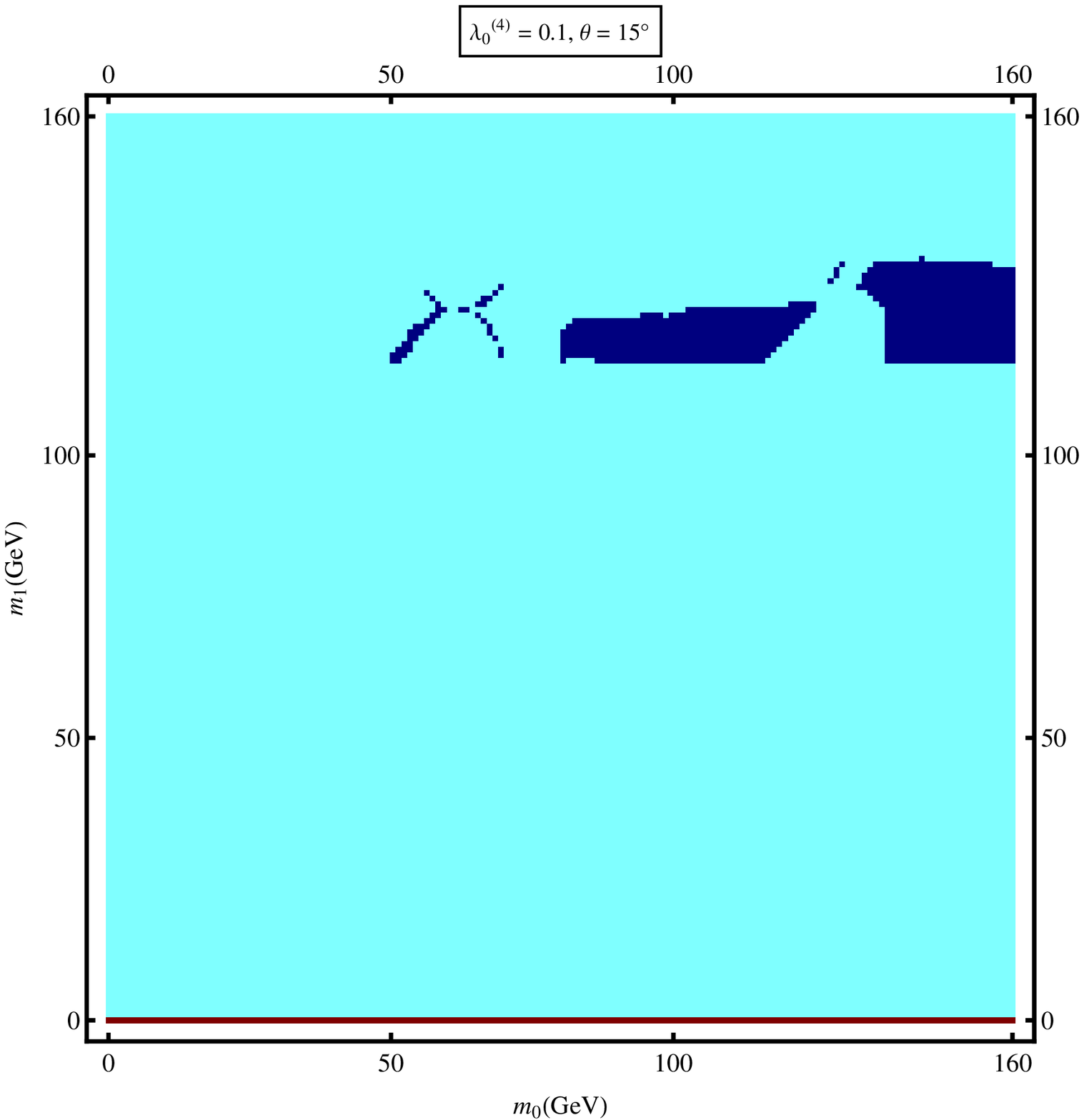}
\caption{The physical Higgs self-coupling $\lambda _{0}^{(4)}$ shrinks the viability region
(blue) as it increases.}
\label{RGE-dens10_lambda04-01_theta-15_tm-6_v2}
\end{figure}

Increasing the Higgs-DM mutual coupling $\lambda _{0}^{\left( 4\right) }$
has the opposite effect, that of shrinking existing viability regions.
Indeed, compare Fig.~\ref{RGE-dens10_lambda04-01_theta-15_tm-6_v2} for which 
$\lambda _{0}^{\left( 4\right) }=0.1$ and $\theta =15^{\mathrm{o}}$ $\left(
\Lambda _{m}\simeq 40\mathrm{TeV}\right) $ with
Fig.~\ref{RGE-dens10_lambda04-001_theta-15_tm-6_v2}. We see shrunk regions, pushed
downward by a few GeVs, which is not a substantial relocation. Remember that
increasing $\lambda _{0}^{\left( 4\right) }$ raises $\eta _{01}\left(
\Lambda _{0}\right) $ well enough above 1 so that this latter will leave the
perturbativity region sooner. Increasing it is also caught up by the
relic-density constraint, which tends to shut down such larger values of
$\lambda _{0}^{\left( 4\right) }$ when the dark-matter mass $m_{0}$ is large.
The direct-detection constraint has also a similar effect.

\section{Concluding remarks}

In this work, we have studied the effects and consequences of the
renormalization group equations at one-loop order on a two-singlet model of
cold dark matter that consists in extending the Standard Model with two real
gauge-singlet scalar fields. The two issues we monitored are perturbativity
and vacuum stability. The former is controlled by the auxiliary-field
self-coupling $\eta _{1}$ and the mutual coupling $\eta _{01}$ between the
dark matter and the auxiliary fields. The latter is controlled by the Higgs
self-coupling $\lambda $. When the non-Higgs SM coupling constants are
switched off, all scalar couplings are positive increasing functions of the
scale $\Lambda $. Reintroducing them flattens the rates for all the scalar
couplings and makes the Higgs coupling $\lambda $ turn negative at some
scale. The mutual couplings $\lambda _{0}$ (DM-Higgs) and $\lambda _{1}$
(Higgs-auxiliary) stay always well below one, whereas $\eta _{01}$ boosts up
for larger $m_{0}$ (DM mass) and/or $m_{1}$ (auxiliary-field mass),
dominating over $\eta _{1}$ in some regions.

We then have investigated the regions in the space of parameters in which
the model is viable. We have plotted these regions in the
($m_{0},m_{1}$)-plane while varying the physical mutual coupling $\lambda _{0}^{(4)}$
between the dark matter $S_{0}$ and the physical Higgs $h$, and the mixing
angle $\theta $ between $h$ and the physical auxiliary field. We have
required that the model reproduces the DM relic density abundance, and that
it complies with the measured direct-detection upper bounds -- those of the
XENON 100 experiment. We have also imposed the RGE perturbativity and
vacuum-stability criteria that we deduced from this work together with a maximum cutoff
$\Lambda _{m}\simeq 40\mathrm{TeV}$, a scale at which heavy degrees of
freedom may start to be relevant, something that could be probed by future
colliders. This analysis has shown that for small $\lambda _{0}^{(4)}$ and 
$\theta $, the auxiliary-field mass $m_{1}$ is confined to the interval
$116\mathrm{GeV}-138\mathrm{GeV}$, while the DM mass $m_{0}$ is confined mainly
to the region above $118\mathrm{GeV}$,\ with a small showing in the narrow
interval $57\mathrm{GeV}-68\mathrm{GeV}$. Increasing $\theta $ enriches the
existing viability regions without relocating them, while increasing$\
\lambda _{0}^{\left( 4\right) }$ has the opposite effect, that of shrinking
them without substantial relocation.

It is pertinent at this stage to comment on the implications of the Higgs
discovery at the LHC on the possibility of having a light dark matter WIMP
$S_{0}$ with a mass $m_{0}\lesssim 62\mathrm{GeV}$, a situation allowed in
this two-singlet model. Indeed, on the one hand, for such a light dark
matter, the decay channel $h\rightarrow S_{0}S_{0}$ becomes open, and
therefore will lower the number of Higgs decays into SM particles. On the
other hand, The ATLAS and CMS published data on Higgs boson searches seem to
indicate that the observed boson is SM-like, and so, one expects to have
stringent constraints on the parameter space when it comes to light
dark-matter masses. In \cite{GKMRS}, a global fit to the Higgs boson data
that includes those presented at the Moriond 2013 conference by the ATLAS 
and CMS  collaborations \cite{ATLAS2013, CMS2013} has been performed; see \cite{GFIT} for earlier fits.
It has been found that any extra invisible Higgs boson decay must be bounded
by the following condition on the corresponding branching ratio: 
\begin{equation}
\mathrm{Br}(h\rightarrow \mathrm{invisible})<19\%.  \label{Br-h-inv}
\end{equation}
It turns out that in our two-singlet model, the branching fraction of the
invisible width of the Higgs boson is smaller than the bound above for
 $m_{0}\lesssim 62\mathrm{GeV}$. Indeed, if we take for example $m_{0} =
55\mathrm{GeV}$ used frequently in this work, the ratio $\Gamma (h\rightarrow
S_{0}S_{0})/{\Gamma (h\rightarrow b{\bar{b}})}$ is less than $17\%$, quite
consistent with the above current bound. Therefore, we conclude that the
two-singlet model is consistent with the current available data regarding
the Higgs boson searches.

Finally, we ask whether the model allows for very light cold dark matter.
Below $5\mathrm{GeV}$, direct detection puts no experimental bound on the
total cross section $\sigma _{\det }$ for non-relativistic elastic
scattering of a dark matter WIMP off a nucleon target. Such a situation
allows for very small $m_{0}$ regions of viability, but only when $\lambda
_{0}^{\left( 4\right) }$ is quite large ($\sim 2$ and above) and $\theta $ not too small
( $\sim 15^{\mathrm{o}}$ and above). However, for such values of the
parameters, the branching fraction of the invisible Higgs decay is larger
than $25\%$, which is excluded by the current LHC available data.


\begin{thebibliography}{99}
\bibitem{ATLAS} G. Aad\textit{\ et al}. [ATLAS Collaboration], Phys. Lett. B 
\textbf{716}, 1 (2012).

\bibitem{CMS} S. Chatrchyan \textit{et al}. [CMS Collaboration], Phys. Lett.
B \textbf{716}, 30 (2012).

\bibitem{dm-observation} P.A.R.~Ade \textit{et al.} [Planck Collaboration], 
{\tt arXiv:1303.5062 [astro-ph.CO]}.

\bibitem{SZ} V. Silveira and A. Zee, Phys. Lett. B \textbf{161}, 136 (1985).

\bibitem{MD} J. McDonald, Phys. Rev. D \textbf{50}, 3637 (1994).

\bibitem{BP} C. P. Burgess, M. Pospelov, and T. ter Veldhuis, Nucl. Phys. B 
\textbf{619}, 709 (2001).

\bibitem{one-singlet} C.~Bird, P.~Jackson, R.~V.~Kowalewski, and
M.~Pospelov, Phys.\ Rev.\ Lett.\ \textbf{93}, 201803 (2004); D.~O'Connell,
M.J.~Ramsey-Musolf, and M.B.~Wise, Phys.\ Rev.\ D \textbf{75}, 037701
(2007); V. Barger, P. Langacker, M. McCaskey, M. J. Ramsey-Musolf, and G.
Shaughnessy, Phys. Rev. D \textbf{77}, 035005 (2008); X.G. He, T. Li, X.Q.
Li, J. Tandean, and H.C. Tsai, Phys. Rev. D \textbf{79}, 023521 (2009); M.
Gonderinger, Y. Li, H. Patel, and M.J. Ramsey-Musolf, JHEP \textbf{01}
(2010) 053; S.~Andreas, C.~Arina, T.~Hambye, F-S.~Ling, and M.H.G.~Tytgat,
Phys.\ Rev.\ D \textbf{82}, 043522 (2010); Y. Cai, X. G. He, and B. Ren,
Phys. Rev. D \textbf{83}, 083524 (2011); J.M.~Cline, K.~Kainulainen, JCAP 
\textbf{1301}, 012 (2013).

\bibitem{Hinv} A.~Djouadi, O.~Lebedev, Y.~Mambrini, and J.~Quevillon, Phys.\
Lett.\ B \textbf{709}, 65 (2012).

\bibitem{XENON10} J.~Angle \textit{et al}. [XENON Collaboration],
Phys.~Rev.~Lett.~\textbf{100}, 021303 (2008).

\bibitem{CDMSII} Z.~Ahmed \textit{et al}. [CDMS Collaboration],
Phys.~Rev.~Lett.~\textbf{102}, 011301 (2009); Science \textbf{327}, 1619 (2010).

\bibitem{XENON100} E.~Aprile \textit{et al.} [XENON100 Collaboration],
Phys.\ Rev.\ Lett.\ \textbf{109}, 181301 (2012).

\bibitem{complex} G.~Belanger, B.~Dumont, U.~Ellwanger, J.~F.~Gunion, and
S.~Kraml, \texttt{arXiv:1302.5694 [hep-ph]}.

\bibitem{one-singlet-difficulties} See also M.~Asano and R.~Kitano, Phys.\
Rev.\ D \textbf{81}, 054506 (2010); Y.~Mambrini, Phys.\ Rev.\ D \textbf{84},
115017 (2011); M.~Farina, M.~Kadastik, D.~Pappadopulo, J.~Pata, M.~Raidal,
and A.~Strumia, Nucl.\ Phys.\ B \textbf{853}, 607 (2011); I.~Low,
P.~Schwaller, G.~Shaughnessy,\ and C.E.M.~Wagner, Phys.\ Rev.\ D \textbf{85},
015009 (2012); Y.~Mambrini, J.\ Phys.\ Conf.\ Ser.\ \textbf{375}, 012045
(2012).

\bibitem{abada-ghaffor-nasri} A. Abada, D. Ghaffor, and S. Nasri, Phys. Rev.
D \textbf{83}, 095021 (2011).

\bibitem{abada-nasri-1} A.~Abada and S.~Nasri, Phys. Rev. D \textbf{85},
075009 (2012).
  
\bibitem{GLR} M.~Gonderinger, H.~Lim, and M.J.~Ramsey-Musolf, Phys.\ Rev.\ D 
\textbf{86}, 043511 (2012).

\bibitem{SM-results} C.~Ford, D.R.T.~Jones, P.W.~Stephenson, and
M.B.~Einhorn, Nucl.\ Phys.\ B \textbf{395}, 17 (1993); M.~Sher, Phys.\
Rept.\ \textbf{179}, 273 (1989); A.~Djouadi, Phys.\ Rept.\ \textbf{457}, 1
(2008).

\bibitem{Planck} P.A.R.~Ade {\it et al.}  [Planck Collaboration],
{\tt arXiv:1303.5076 [astro-ph.CO]}.

\bibitem{GKMRS} P.P.~Giardino, K.~Kannike, I.~Masina, M.~Raidal, and
A.~Strumia, \texttt{arXiv:1303.3570 [hep-ph]}.

\bibitem{ATLAS2013} V. Martin,  Talk presented at Rencontres de Moriond - 2-16 March 2013, La Thuile.

\bibitem{CMS2013} C. Ochando, Talk  presented at Rencontres de Moriond - 2-16 March 2013, La Thuile.

\bibitem{GFIT} G.~Belanger, B.~Dumont, U.~Ellwanger, J.~F.~Gunion,\ and
S.~Kraml, \texttt{arXiv:1302.5694 [hep-ph]}; P.P.~Giardino, K.~Kannike,
M.~Raidal, and A.~Strumia, JHEP \textbf{1206}, 117 (2012); J.R.~Espinosa,
M.~Muhlleitner, C.~Grojean, and M.~Trott, JHEP \textbf{1209}, 126 (2012).
\end{thebibliography}
\end{document}